\documentstyle[epsf]{mn2e}

\DeclareMathAlphabet{\mathbi}{\encodingdefault}{\rmdefault}{\bfdefault}{\itdefault}
\DeclareRobustCommand{\bit}[1]{\ifmmode\mathbi{#1}\else\textbf{\textit{#1}}\fi}

\newcommand{\be}{\begin{equation}}
\newcommand{\ee}{\end{equation}}
\newcommand{\Mp}{m_{\rm p}}

\title[SZE in WMAP]{The Sunyaev-Zel'dovich effect in WMAP data.} 
  
\author[Diego \& Partridge]  
  {J.M Diego$^{1}$, B. Partridge$^{2}$\\  
   $^1$ IFCA, Instituto de F\'\i sica de Cantabria (UC-CSIC). Avda. Los Castros 
s/n. 39005 Santander, Spain.\\
   $^2$ Department of Astronomy, Haverford College, Haverford, PA, 19041, USA.}
\date{Draft version \today}

\pagerange{\pageref{firstpage}--\pageref{lastpage}}  
  
\begin{document}  
\maketitle  
  
\label{firstpage}  
\begin{abstract}  
Using WMAP 5 year data, we look for the average Sunyaev-Zel'dovich effect (SZE) signal from clusters of galaxies by stacking the regions around hundreds 
of known X-ray clusters. We detect the average SZE at a very high significance level. The average cluster signal is 
spatially resolved in the $W$ band. This mean signal is compared with the expected signal from the same clusters calculated on the basis of 
archival ROSAT data. 
From the comparison we conclude that the observed SZE seems to be less than the expected signal derived from X-ray measurements when 
a standard $\beta$-model is assumed for the gas distribution. This conclusion is model dependent. 
Our predictions depend mostly on the assumptions made about the core radius of clusters and the slope of the gas density profile. 
Models with steeper profiles are able to simultaneously fit both X-ray and WMAP data better than a $\beta$-model. 
However, the agreement is not perfect and we find that it is still difficult to make the X-ray and SZE results  agree. 
A model assuming point source contamination in SZE clusters renders a better fit to the one-dimensional SZE  
profiles thus suggesting that contamination from point sources could be contributing to a diminution of the SZE signal. 
Selecting a model that better fits both X-ray and WMAP data away from the very central region, we estimate the level of contamination and find that on average, 
the point source contamination is on the level of 16 mJy (at 41 GHz), 26 mJy (at 61 GHz) and 18 mJy (at 94 GHz). These 
estimated fluxes are marginally consistent with the estimated contamination derived from radio and infrared surveys thus 
suggesting that the combination of a steeper gas profile and the contribution from point sources allows us  to consistently explain the 
X-ray emission and SZE in galaxy clusters as measured by ROSAT and WMAP.
  
\end{abstract}  

\section{introduction}\label{sec_into}
In the last few years years, there has been an ongoing debate about the possibility that we do not yet quite understand  
the physics involved in the interaction between the gas inside galaxy clusters and the cosmic microwave background (or CMB). 
This interaction between CMB photons and galaxy clusters is described 
by the Sunyaev-Zel'dovich effect (or SZE hereafter; Sunyaev \& Zel'dovich 1972). 
The motivation for this debate is the apparent weaker strength of the SZE signal  in galaxy clusters than expected from X-ray observations of the same clusters 
(Lieu et al. 2006, Afshordi et al. 2007, Atrio-Barandela et al. 2008). 
On the other hand, other works claim that there is no missing signal and that the interaction between galaxy 
clusters and the CMB can be well described in terms of the SZE once a realistic model for the gas distribution is considered (Mroczkowski et al. 2009).
Understanding the SZE is important also for cosmological studies based on clusters and the SZE 
(see for instance Bartlett \& Silk 1993, Barbosa et al. 1996, Aghanim et al. 1997, Diego et al 2002b, Carlstrom et al. 2002,  Komatsu \& Seljak 2002, Huffenberger et al. 2004). 

In the first attempt to detect the SZE in WMAP data, Diego et al (2003) found no SZE when 
correlating WMAP one year maps and maps derived from the ROSAT All Sky Survey 
(or RASS hereafter, Snowden et al. 1997). In this work the authors 
were not selecting only areas of the sky with known clusters; rather a whole section of the sky was used in the analysis. In more recent years, 
Atrio-Barandela et al. (2008) used a large sample of know galaxy clusters and observed that while the observed signal at
small radii could be consistent with the expected SZE signal from the central parts of the clusters, the signal at larger radii 
(1 degree or more) is not consistent with the predictions. 
In the same work it was suggested that the pressure profile follows a Navarro-Frenk-White (or NFW, Navarro et al. 1996) 
profile rather than a $\beta$-model profile. 
This result agrees with higher resolution X-ray observations where profiles steeper than a standard $\beta$-model 
seem more adequate to fit the observations (Neumann 2005, Vikhlinin et al. 2005, 2006, Croston 2008) and also with theoretical predictions (Hallman et al. 2007). In a different work, Myers et al. (2004) 
used the APM Galaxy Survey and the Abell-Corwin-Ollowin (ACO) catalogue and through a correlation analysis with WMAP data,  
found evidence that significant amounts of gas lay in the outskirts of galaxy clusters (at distances up to 5 Mpc). 
Hern\'andez-Monteagudo \& Rubino-Martin (2004) used an X-ray catalog of clusters to cross-correlate with WMAP first year data. 
They found a 2-5 $\sigma$ correlation signal and a mean temperature decrement of 15-35 $\mu K$ in the Rayleigh-Jeans region. 
However, no comparison was made with the expected SZE signal from these clusters. 
One of the most closely related works to the present paper can be found in Lieu et al (2006) where they use a sample of 
31 X-ray clusters and look for the SZE in WMAP one year data. They found evidence that there is less SZE than expected. The strength 
of their results is based on the fact that the models they used to predict the SZE are consistent with the X-ray emission observed for 
each one of the 31 clusters. Afshordi et al. (2007) also found evidence of missing SZE signal using a sample of 193 clusters.
Our work will have some similarities with Lieu et al. (2006) but also some important differences, 
for instance the much larger size of our cluster sample. Also, we consider different models which could explain some of the differences 
between the results. The reader will also find some similarities with the work presented in Atrio-Barandela 
et al. (2008), in particular the cluster sample used in both papers is very similar but the analysis (and conclusions) 
differ significantly. 

In this paper we follow previous work and revisit the debate, but also perform a different analysis 
using the most accurate CMB data to date, namely the five year WMAP data (Hinsaw et al. 2009). 
We use a large sample of known X-ray clusters and stack the CMB data in an area around each one 
of them. The stacking procedure effectively reduces the contributions of instrumental noise and fluctuations in the CMB itself relative 
to the SZE making it possible to {\it see} the clusters (or their average signal) in WMAP data. For this process, it is crucial to have 
a large sample of clusters because otherwise the noise, and especially the CMB fluctuations, might not average to zero. 
Using the known positions of galaxy clusters offers the unique advantage of knowing were to look for the SZE. 
But  it also provides us with a way of modeling the expected signal. Using the X-ray fluxes we are able to predict the SZE for each cluster 
(and hence, the average signal when the SZE signal of all clusters is added together) and compare the average predicted SZE with the observations. 
We find that the expected SZE signal depends strongly on the assumptions made about the internal distribution of the cluster gas. 
Our results neither strongly contradict nor strongly confirm the findings in some previous works where a lack of signal from 
clusters at mm wavelengths was found. 
As we will show, our results still show a lack of SZE signal in galaxy clusters but the disagreement between the observed signal 
and the expected one is significantly less than previously reported by Lieu et al (2006). 

We also explore some of the possible mechanisms responsible for the apparently {\it missing} SZE signal. 
One of the most likely explanations for the lack of SZE signal is (apart from the internal distribution of the gas) 
contamination from radio and/or infrared galaxies inside the cluster. 
Since the SZE signal is negative at the frequencies used by WMAP and the radio and infrared emission is positive, a reduction of the SZE 
signal might occur from this contamination. Lin et al. (2009) made a survey of radio galaxies inside galaxy clusters in the frequency 
range 4.9-43 GHz. They found that half the observed sources had a step spectrum (as expected) but they also found that most of the 
unresolved sources had flat or even inverted spectrum. These flat and inverted spectrum sources have more flux at higher frequencies 
than predicted from standard extrapolations (that usually assume a steep spectra) and they could be a significant source 
of contamination even at WMAP frequencies. On the other extreme of the frequency, SCUBA observations of galaxy clusters (Zemcov et al. 2007) 
also reveal a significant amount of infrared emission in galaxy clusters at SCUBA frequencies. We examine both in section \ref{section_PS}.  
We discuss there whether the point sources are or are not significantly reducing the amount of SZE signal in WMAP.
 
\section{Sunyaev-Zel'dovich Effect and X-rays in clusters}\label{section_1}  
When CMB photons cross a galaxy cluster there is a probability that some of the photons interact with the free electrons in the 
hot plasma through inverse Compton scattering. 
Through this interaction, the CMB photons can gain a small amount of energy. This change of energy distorts the blackbody 
spectrum of the CMB photons. At frequencies smaller than 217 GHz the distortion results in a deficit of photons compared  with the 
undistorted blackbody spectrum. These {\it missing} photons appear at (or are promoted to)  higher frequencies creating an excess in flux 
above 217 GHz. This is known as the Sunyaev-Zel'dovich effect (Sunyaev \& Zel'dovich 1972). 
At WMAP frequencies, the clusters will appear colder than the CMB since 
we are in the frequency range where the deficit occurs ($\nu < 217$ GHz). 
The temperature decrement (increment for frequencies $\nu > 217$ GHz) observed in the direction $\theta$ can be 
described as;
\begin{equation}
\Delta T(\theta) = T_o \int n_e T dl 
\label{eq_DeltaT}
\end{equation}
where $T_o$ contains all the relevant constants including the frequency dependence ($g_x = x(e^x + 1)/(e^x - 1) - 4$ with $x=h\nu/kT$), 
$n_e$ is the electron density and $T$ is the electron temperature. The integral 
is performed along the line of sight.
An interesting quantity is the integrated temperature decrement which can be obtained after integrating 
the above equation over the solid angle of the cluster ($d\Omega$)
\begin{equation}
\Delta T_{Tot} \propto \int n_e T d\Omega dl \propto \frac{M_{gas}T}{D_a(z)^2}
\label{Eq_2}
\end{equation}
where $M_{gas}$ is the total gas mass of the cluster and $D_a(z)$ the angular diameter distance 
(instead of the gas mass one can use the total cluster mass if $M_{gas} = f_b*M$ and the baryon fraction, $f_b$, is constant). 
The advantage of using this expression is that the total integrated signal is independent of the internal distribution of the gas. 
Once the mass, temperature and redshift of the cluster are know, a robust estimate of the integrated $\Delta T_{Tot}$ can be derived. 
The temperature can be obtained from X-ray observations of galaxy clusters. 
The mass can also be derived from the same observations using scaling relations between the X-ray luminosity and the total mass. 
The total SZE flux is determined from equation (\ref{Eq_2}) and the appropriate frequency dependence conversion between brightness 
and temperature.  
The above expressions describe the thermal SZE. If the gas in the cluster is moving with a significant bulk velocity with 
respect to the CMB, another effect, the kinetic SZE, occurs in addition to the thermal SZE. In this work we will not consider the 
kinetic effect since we will be working with the average signal of a large number of clusters, and in this case the average bulk 
velocity can be neglected. 

The same electrons that interact with the CMB photons emit energy in the form of X-rays through the bremstrahlung process. 
An interesting situation is when the plasma has very high temperatures ($T > 10$ keV). 
In this case, a non-negligible fraction of the electrons have relativistic velocities (and energies) and 
are able to promote some of the CMB photons into the X-ray band itself. This mechanism increases the amount of observed X-rays and reduces the amount 
of CMB photons. This interaction has been suggested in the past as a possible explanation for the reduced SZE predicted from X-ray observations 
(Lieu \& Quenby 2006). However, few of the clusters we consider have $T > 10$ keV. 
Considering only the non-relativistic case, the observed X-ray flux in the same direction $\theta$ from a cluster is given by;
\begin{equation}
S_x(\theta) = S_o \frac{\int n_e^2 T^{1/2} dl}{D_l(z)^2} 
\label{eq_Sx}
\end{equation}
where $D_l(z)$ is the luminosity distance. The quantity $S_o$ contains all the relevant constants and corrections (including the gaunt factor, 
band correction and k-correction). 

\begin{figure}  
   \epsfysize=6.0cm   
   \begin{minipage}{\epsfysize}\epsffile{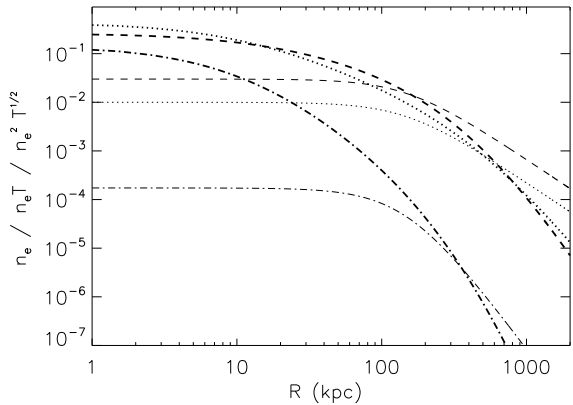}\end{minipage}  
   \caption{$\beta$-model (thin lines) compared with AD08 model (thick lines). In both cases, the dotted 
            line represents the electron density ($n_e$), the dashed line the electron pressure  ($n_e T$ or 
            SZE) and the dotted-dashed line the X-ray emissivity ($n_e^2 T^{0.5}$). The $\beta$-model 
            corresponds to an isothermal model with $T=3$ keV, $\beta=2/3$, $r_c = 150$ kpc, 
            and $n_o = 0.01$ cm$^{-3}$. The AD08 model has $n_o = 0.01$ cm$^{-3}$, $T_0=3$ keV, 
            $a = 3r_c = 450$ kpc, $t=0.2$, $\alpha=0.15$ and $n=4$. The AD08 model is explained 
            in the text.}
   \label{fig_BetaYago}  
\end{figure} 

An obvious advantage of combining X-ray and SZE observations is that it is in principle possible to break the degeneracy between the multiple 
density profiles due to the difference dependence of the two signals on $n_e$.  

One of the most common assumptions made when modeling the gas distribution in galaxy clusters is to assume 
that $n_e$ is well described by a $\beta$-model. 
\begin{equation}
n_e(r) = \frac{n_o}{[1 + (r/r_c)^2]^{3\beta/2}}
\label{eq_beta}
\end{equation}
The $\beta$-model is well motivated by observations of galaxy clusters and theoretical predictions. The exponent $\beta$ typically ranges from values of 
0.5 to 0.9 but for practical reasons, a value of $\beta = 2/3$ is usually assumed (this choice allows for an analytical solution 
of the integral of $n_e$). Other more elaborate (and accurate) models can be used instead. As an alternative model we will consider the following described  
in Ascasibar \& Diego (2008). We will refer to this model as AD08 from now on. Ascasibar \& Diego (2008) have shown how this model 
is capable of fitting high resolution profiles of X-ray clusters obtained with CHANDRA data (Vikhlinin et al. 2006). 
The AD08 model describes the radial structure of clusters in terms of five free parameters: the total mass $M$ of 
the system (or the characteristic temperature $T_0$), a characteristic radius $a$, 
the cooling radius in units of $a$, $0<\alpha<1$, the central temperature in units 
of $T_0$, $0<t<1$, and the asymptotic baryon fraction at large radii, $f\sim1$.                                                                        
Ascasibar \& Diego (2008), have shown that this model is able to reproduce 
the observed X-ray properties of real clusters. 
The total (gas plus dark matter) density follows a Hernquist (1990) profile,                                                                
\begin{equation}                                                                
\rho(r)=\frac{M}{2\pi a^3}\frac{1}{r/a(1+r/a)^3},                               
\label{eqRhoH}                                                                  
\end{equation}                                                                  
whereas the gas temperature varies according to                                 
\begin{equation}                                                                
T(r)=\frac{T_0}{1+r/a}~\frac{t+r/a_{\rm c}}{1+r/a_{\rm c}}.                     
\label{eq_Yago2}
\end{equation}       

From hydrostatic equilibrium the mass-temperature relation can be obtained :              
\begin{equation}                                                                
(n+1)\frac{kT_0}{\mu\Mp}=\frac{GM}{a}                                           
\label{eqMT}                                                                    
\end{equation}                                                                  
and the gas density profile is                                                     
\begin{equation}                                                                
\frac{n_e(r)}{n_o} =                                              
  \left( \frac{1+r/a}{t\alpha+r/a} \right)^{1+\frac{\alpha-t\alpha}{1-t\alpha}(n+1)}   
         \frac{\alpha+r/a}{\left(1+r/a\right)^{n+1}}       
\label{eq_Yago1}
\end{equation}   
where $n$ is an effective polytropic index, $k$ is the Boltzmann constant, $\Mp$ 
denotes the proton mass, $\mu\simeq0.6$ is the molecular weight of the gas, and 
$n_o$ is the central gas density. 
As in AD08, we set $n=4$ in order to obtain a constant baryon fraction at large 
radii. From Ascasibar \& Diego (2008) we derive average values for  
the $t$ and $\alpha$ parameters of the model. in particular we take $t=0.2$ and 
$\alpha = 0.15$. For the parameter $a$ we use $a = 3.3r_c$ which makes the AD08 model 
resemble a $\beta$-model with the same $r_c$ at intermediate radii. 

A comparison of the density profiles of the two models is shown in figure \ref{fig_BetaYago}. In the same figure we also 
show the SZE and X-ray emissivity for the two models. In our calculations, we assume that the $\beta$-model is 
isothermal ($T=const.$). The AD08 model is steeper than the $\beta$-model. This has important implications since the gas is more 
concentrated in the center and therefore less gas is needed in order to explain a given X-ray flux. Also, clumpiness inside the cluster can 
create more X-rays with the same amount of gas due to the same mechanism (since the X-ray emissivity varies as $n_e^2$). 
For instance, one can assume that some amount of clumpiness exist around the galaxies belonging to the cluster. 
High resolution observations with CHANDRA have reveled departures from the 
smooth and continuous X-ray emissivity assumed in the construction of figure \ref{fig_BetaYago}. An example are the so called 
cold fronts first observed by CHANDRA (see for instance Markevitch et al. 2000, 
Vikhlinin et al. 2001, Mazzotta et al. 2001 and others). 
In the X-ray band, the cold fronts appear as a shock wave with 
significantly more surface brightness inside the shock wave than outside. The interpretation of these shock waves is that they 
are produced by the oscillations of the gaseous core around the peak of gravitational potential. 
When the gas oscillates and moves to a region with lower potential energy, it expands and cools. 
Then, because it is cooler, it collapses to higher density, thus increasing $n_e^2$ and the X-ray luminosity 
(Mathis et al. 2002, Nagai \& Kravtsov 2003, Ascasibar \& Markevitch 2006). 
Studies have also revealed that the pressure remains constant 
across this shock wave (e.g. Mazzotta et al. 2001, Nagai \& Kravtsov 2003). 
This means that while the X-ray brightness can be enhanced by mechanisms like 
the cold fronts, the SZE remains constant, since the pressure does not change. 
We return to this issue in section \ref{section_discussion}

\section{X-ray cluster catalogs}

\begin{figure}  
   \epsfysize=6.0cm   
   \begin{minipage}{\epsfysize}\epsffile{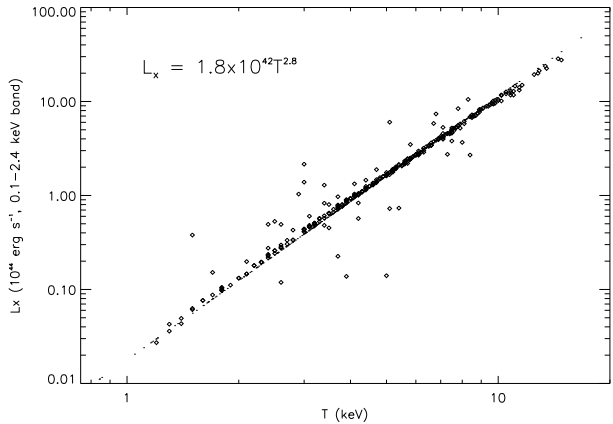}\end{minipage}  
   \caption{Luminosity temperature relation. The diamonds correspond to the BCS and eBCS catalog while 
            the dots correspond to the REFLEX catalog.}
   \label{fig_LT}  
\end{figure} 

There are a variety of cluster catalogs. For the purpose of this work the most interesting ones are those based on an 
X-ray selection. There are two important advantages when using a cluster catalog to look for the SZE. 
The first one is obvious: knowing the positions of galaxy clusters 
allows us to select those portions of the sky where a cluster exists. Second, as mentioned earlier, X-rays 
are emitted by the same hot gas that produces the SZE. The X-ray signal of a galaxy cluster then allows to 
constrain the model used to describe the internal gas distribution. 

For this work, we build a large cluster catalog combining three catalogs: the Bright Cluster Sample 
(or BCS Ebeling et al. 1998), the extended Bright Cluster Sample (or eBCS Ebeling et al. 2000) and the 
ROSAT-ESO Flux-Limited X-Ray catalog (or REFLEX; B\"ohringer et al. 2004). These catalogs are X-ray selected using data from 
the ROSAT All Sky Survey (or RASS). The combined catalog covers both the north and south Galactic 
hemispheres for Galactic latitudes roughly above 20$^{\circ}$ and below -20$^{\circ}$. There are 750 clusters 
in our combined catalog outside the Galactic exclusion zones. 
The combined catalog contains measured redshifts, X-ray fluxes, temperatures (for some clusters) and  luminosities for the brightest 
X-ray clusters observed with ROSAT. 
In the case of REFLEX clusters, the catalog does not contain temperature estimates, but these can be derived from scaling relations as described 
below. A scaling relation was also used by Ebeling et al. (2000) to derive temperatures for some of the clusters in the eBCS catalog 
(see figure \ref{fig_LT}). 

Any model that we use to predict the SZE in clusters can be compared with the observed X-ray emission in those clusters. 
In an ideal scenario, we would like to do this exercise on a one-by-one basis; Fiting a model to the X-ray surface brightness 
of a particular cluster and using the same model to predict the amount of SZE in that cluster, then repeating the process for every 
single cluster. This kind of exercises has been done in the past (Lieu et al. 2006) but with a small number of very bright clusters. 
In section \ref{section_discussion} we will discuss in more detail the interesting findings of Lieu et al. (2006). 
Only very bright clusters allow a reliable estimation of cluster profiles with ROSAT data. A small (but growing) number of clusters have 
been observed with CHANDRA and XMM (Vikhlinin et al. 2006, Croston et al. 2008). 
These observations allow a much more detailed analysis. 

In this work we adopt a different approach in which the sample size matters. 
We focus on having a large sample of clusters (even if their profiles are not resolved) 
rather than a small but well defined sample of clusters. Instead of fitting the profile of a few dozen clusters we fit the average flux of a much wider 
sample of hundreds of clusters. 
Taking a large sample of clusters is crucial for reducing the CMB and instrumental noise fluctuations 
that contaminate the SZE signal. We use the X-ray observed flux in every cluster to {\it normalize} the gas model. By normalizing 
we mean selecting the particular value of $n_o$ in equations (\ref{eq_beta}) and (\ref{eq_Yago1}) such that the total flux predicted for our model 
coincides with the observed flux. Of course, this alone does not guarantee that one-by-one we are achieving a good fit to the surface 
brightness of every cluster. We are more interested in the {\it average} properties of clusters. For instance, most clusters present 
significant departures from circular symmetry but the average profile of a few clusters already starts to show this symmetry.

\begin{figure}  
   \epsfysize=6.0cm   
   \begin{minipage}{\epsfysize}\epsffile{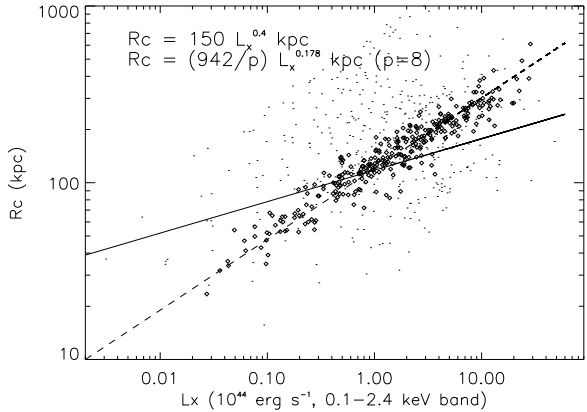}\end{minipage}  
   \caption{Relation between the X-ray luminosities and the core radius. The diamonds represent the BCS and eBCS VTP radius 
            (see text) divided by a factor 3 and the dots represent the rough estimate of the core radius given in the REFLEX catalog. 
             The solid line shows the theoretical relation expected for a self-similar model while the dashed line represents and alternative 
             and arbitrary model with slope 0.4. }
   \label{fig_RL}  
\end{figure}

Average properties can usually be described by scaling relations, and are powerful tools that have allowed us to set constraints on 
cosmological models using galaxy cluster observations (Oukbir \& Blanchard 1992, Lupino \& Gioia 1995, 
Donahue 1996, Eke et al. 1996, Kitayama \& Suto 1997, Oukbir \& Blanchard 1997, Mathiesen \& Evrard 1998, Donahue \& Voit 1999, 
Diego et al. 2001, Henry 2004). Some of these scaling relations show very tight 
correlations between cluster properties. The best studied so far use X-ray data. In particular, one of the most famous is the correlation 
between the  X-ray luminosity ($L_x$) and the plasma temperature ($T$) or $L_x-T$ relation (David et al. 1993, Mushotzky \& Scharf 1997, 
Vikhlinin et al. 2002, Ettori et al. 2004). 
N-body simulations, as well as analytical 
models, predict similar correlations between quantities derived from SZE observations. Scaling relations are representative of the average properties 
of clusters and we use them to model the average signal observed by two very different experiments, ROSAT and WMAP. Two examples of scaling 
relations used  in this work are shown in figures  \ref{fig_LT}  and  \ref{fig_RL}.  

In figure \ref{fig_LT} we show the $L_x-T$ relation for clusters in the different catalogs. The diamonds show the 
corresponding temperature and X-ray luminosity (in the 0.1-2.4 keV band) for BCS and eBCS clusters while the dots correspond 
to the REFLEX clusters. In the case of BCS and eBCS clusters, the X-ray luminosities were given originally for a different cosmological 
model than the one used in the REFLEX catalog. Starting from the fluxes, we compute the luminosities for BCS and eBCS clusters but for 
the same cosmological model used in the REFLEX catalog. Some of the clusters in the BCS and eBCS catalogs did not have temperature estimates. 
In these cases, the temperature was obtained from the luminosity using a scaling relation. In the case of REFLEX clusters, none of the clusters 
in the catalog had temperature estimates. We used a scaling relation similar to the one in the BCS and eBCS clusters. 
\begin{equation}
L_x = 1.8 \times 10^{42} T^{2.8} \ \ erg/s,
\end{equation}
where $L_x$ is the luminosity in the $0.1-2.4$ keV band and T is the gas temperature in keV. 

A relation between the X-ray luminosity and physical core size of the cluster can be obtained from the above $L_x-T$ relation 
and the following two scaling relations for a model that we will refer as self-similar model.    
These are the relation between the virialized mass and temperature (Horner \& Mushotzky 1999, Diego et al. 2001)
\begin{equation}
T \approx 8 M^{2/3} \ \ keV
\end{equation}
and a relation between the virialized mass and the virial radius. (Horner \& Mushotzky 1999, Diego et al. 2001) 
\begin{equation}
r_v \approx 1.3 M^{1/3} (1+z)^{-1}  \ \ Mpc
\end{equation}
In the above expressions, $M$ is the mass in units of $10^{15} M_{\odot}$. If we assume a fixed ratio of the virial to core radius 
$r_v/r_c = p$ then we finally obtain.
\begin{equation}
r_c = \frac{942}{p(1+z)} L_x^{0.178} \ \ kpc
\end{equation}
where $L_x$ is the luminosity in the $0.1-2.4$ keV band in units of $10^{44} \ erg/s$. This scaling law is shown in figure 
\ref{fig_RL} (solid line) and compared with a rough estimate of the core radius for REFLEX clusters (dots) 
as given by Bohringer et al. (2004).  In the case of the BCS
and eBCS clusters, we plot the Voronoi Tesellation Percolation (VTP)
radius, divided by an arbitrary factor of three to omake the BCS+eBCS radii 
consistent with the REFLEX estimates. 
The VTP radius does not have a physical meaning but rather is the 
radius at which the cluster is seen as an overdense region 
in the ROSAT PSPC detector with the VTP algorithm (isophotal radius). 
The diamonds in figure \ref{fig_RL} should not be 
confused with the core radius but rather with a very rough estimate of the physical size of the 
core region in the BCS and eBCS clusters. 
Scaling relations similar to the one shown in figure \ref{fig_RL} have been found in the 
past between the isophotal size and the X-ray temperature (Mohr et al. 2000). 
Also shown in the same figure is a different (and arbitrary) model where
the core radius scales as $L_x^{0.4}$. We will use this arbitrary model as a way to test significant 
departures from the self-similar model. 
Observationally, the radius-luminosity relation is not as well established as the 
$L_x-T$ relation. As we will see later, this will be probably the most important limiting factor on our conclusions as the 
predicted signal depends strongly on the core radius. In particular, in both cases, X-rays and SZE, the total integrated 
flux scales as $r_c^3$. This means that a factor 2 difference in the scaling relations leads to an order magnitude difference 
in the predicted flux. The good news is that the combination of X-ray and SZE measurements helps reduce the range of 
possible models. 

An interesting feature of our normalization process 
is that since we require that the X-ray flux from our model match the observed flux, smaller core radii will boost the central 
electron density (which enters like $n_o^2$ in the predicted X-ray flux) in order to compensate for the $r_c^3$ decrease. 
As a consequence the predicted cluster signal will appear more concentrated in the center but with a large amplitude relative 
to models with larger core radii. Since the X-ray tails of the gas density are difficult to see (or detected by an X-ray experiment) 
the cluster will look as if it is brighter when taking smaller core radii (this is of course a visual artifact as the total flux is the 
same for all models given a particular cluster). In contrast, the predicted flux in the SZE case is not constrained, and 
can be larger or smaller depending on the model. 
It is instructive to look at the order zero expression for the expected total fluxes. In the case of the SZE, the integrated total 
flux is proportional to $n_o \times r_c^3$ (times a slow varying function which depends on the ratio $p=r_v/r_c$). 
On the other hand, the total X-ray flux goes like $n_o^2 \times r_c^3$ (again times another geometric factor 
with a weak dependence on $p$). Since the total X-ray flux is constrained to a value, $S_{obs}$, the electron density is just 
$n_o = S_{obs}^{1/2}\times r_c^{-3/2}$. Substitution of $n_o$ into the SZE flux results in the SZE having a total flux going like 
$S_{obs}^{1/2} \times r_c^{3/2}$ for a given cluster. That is, while the total X-ray flux is constrained to be equal to $S_{obs}$, 
the total predicted SZE flux is not constrained and grows with the core radius as $r_c^{3/2}$.

\section{Stacking of astrophysical data}
When looking at individual clusters in the WMAP data, their signal to noise 
ratio is too small to allow a clear detection of most individual clusters. 
The noise in this case is a combination of the instrumental noise 
(well approximated to first order as a random Gaussian noise) plus  
the CMB. At a lower level, Galactic foregrounds and possibly unresolved 
extragalactic sources (i.e radio and/or infrared galaxies) also contribute 
to the noise fluctuations. One way to 
increase the signal to noise ratio for galaxy clusters is by stacking 
the area around known galaxy clusters (observed for instance by their
X-ray emission). The stacking procedure reduces the contribution of the 
CMB and instrumental noise since these two signals
should average to zero for a sufficiently large number of stacked fields. 
In the case of the Galactic foregrounds and extragalactic unresolved sources, 
these two components do not average to zero in the stacking process, but will
contribute as a largely uniform background to the average map (especially if one 
considers areas of the sky far enough from the Galactic plane). 

\begin{figure}  
   \epsfysize=8.0cm   
   \begin{minipage}{\epsfysize}\epsffile{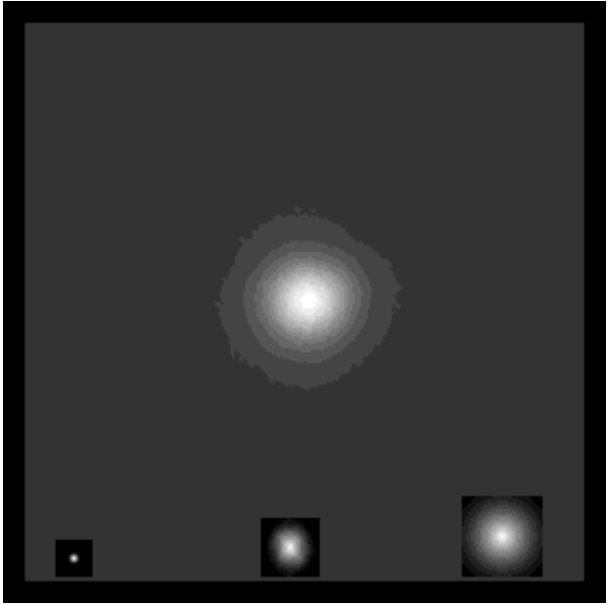}\end{minipage}  
   \caption{Average cluster profile as derived from RASS maps.
            The fields of view in this figure and figure 6 is $2.51^{\circ}\times2.51^{\circ}$ 
            and both are centered on the clusters in the X-ray catalog. 
	    Also shown (on the same scale) are the PSF of the ROSAT PSPC detector (bottom left), 
            the PSF introduced by the {\small HEALPIX} pixelization (bottom middle) and the PSF due 
            to RASS 12 arcminute  pixelization (bottom right). }
   \label{fig_RASS_2D}  
\end{figure}

In our stacking procedure we first construct full sky maps of the data to be stacked
(in those cases where the all sky maps are not already available). An area (or field of view,
FOV hereafter) of $2.5^{\circ}\times 2.5^{\circ}$ is then selected 
around each galaxy cluster in our combined catalog and the area is 
projected into a plane using an orthographic projection. We sample the 
FOV of 2.5 degrees with $256\times256$ pixels (i.e a pixel size of $\approx 0.6$ arcmin). 
Due to the random orientations introduced by the stacking procedure, the 
original pixels in the all-sky maps take random orientations and do not align 
with other pixels in different FOVs. The stacking procedure introduces an average pixel resulting from 
the average of all the possible orientations of the original pixel. In the next section we will discuss how 
to estimate this effect quantitatively, using the X-ray data as an example.

\begin{figure}  
   \epsfysize=6.0cm   
   \begin{minipage}{\epsfysize}\epsffile{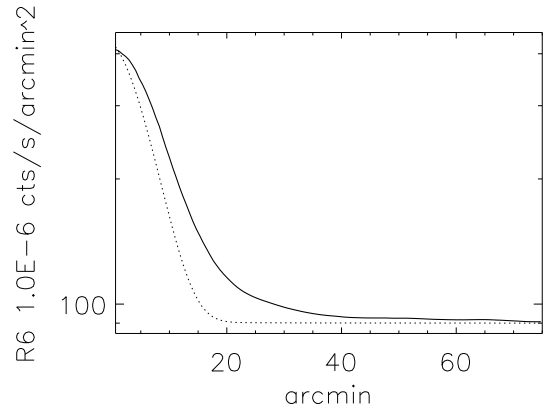}\end{minipage}  
   \caption{One dimensional profile of RASS stacked clusters (solid line) compared with 
            the normalized effective point spread function (or psf, dotted line). 
            This psf contains the combined effect
	    of the intrinsic psf of the ROSAT pspc detector, the 12 arcminute RASS pixelization 
            and the $Nside=1024$ {\small HEALPIX} pixelization. }

   \label{fig_RASS_1D}  
\end{figure}

\section{X-ray Data/ROSAT Data}\label{section_stack}
We use the diffuse X-ray maps derived from the ROSAT All Sky Survey (RASS) maps. These 
maps are described in detail in Snowden et al. (1997). 
The diffuse X-ray RASS maps are given in six Zenithal Equal Area (or ZEA) projections and with a pixel 
size of 12 arcminutes. 

The brightest point sources have been removed using
the full resolution images of the PSPC detector. The extraction radius for these 
sources is at least 3.5 arcminutes. As a consequence of the bright source removal, the 
central peak in some bright clusters is also removed. 


The bright source catalog used in Snowden et al. (1997) to mask out point sources included at least 
400 cluster candidates (see a more detailed description of the bright source catalog in Voges et al. 1999). 
Some of those candidates correspond to real clusters and some not.
We have cross-correlated our catalog containing 750 clusters and the bright source catalog of Voges et al 
(1999) and found that 314 of the objects in the Voges et al. (1999) bright source catalog satisfy the criteria of being 
at an angular distance of less than 1 arcminute from one of the clusters in our sample and have an extent of 30 
arcseconds or more. Based on this estimation we can say that we expect 40\% of our sample to suffer from masking of their 
central region.  We return to this issue in section 7.

The RASS maps are given in different energy bands. We use the R6 band (0.91-1.31 keV) for which cluster emission
is supposed to be maximal compared with emission from the Galaxy (dominant at lower energies) and the unresolved X-ray
background (more relevant at higher energies). At each band the maps are presented in 6 azimuthal projections.
We combine these 6 projections into a single all-sky map using the {\small HEALPIX} pixelization (G\`orski et al. 2005). The original 12 arcminute RASS
pixels are oversampled using a 3.4 arcminute {\small HEALPIX} pixel (Nside=1024 in {\small HEALPIX}). We also subdivide each RASS pixel into 121
square pixels to avoid missing some of the smaller {\small HEALPIX} pixels. The final {\small HEALPIX} map contains over 12 million pixels
and with an effective resolution which is the combination of 3 smoothing kernels. The first smoothing kernel represents the
intrinsic Point Spread Function (or PSF hereafter) of ROSAT's PSPC detector. This PSF depends on the energy and the off-axis
distance (Hasinger et al. 1994). We assumed an energy of 1 keV (consistent with the energy range of the R6 band) and an off-axis distance of 20
arcminutes which is approximately the average distance of a serendipitous source in the field of view of the PSPC detector.
With these parameters the PSF of ROSAT has a typical scale of 1-2 arcminutes (Hasinger et al. 1994). The second kernel represents the stacking
procedure of the original 12 arcminute RASS pixels. Since the center of a galaxy cluster can fall in any position within
the 12 arcminute RASS pixel (and not just in the center), this has to be taken into account.
The stacking procedure averages all the possible orientations of the 12 arcminute pixel containing the center of each galaxy
cluster. The specific form of this kernel is given below. The third kernel is similar to the second one but it accounts for
the stacking of the 3.8 arcminute {\small HEALPIX} pixels. Both, the second and the third smoothing kernel can be described by the
following formula
\begin{equation}
K_{pix}(\theta) = (1 - \theta/\theta_{max})^2 \ \ if \ \ \theta < \theta_{max} \ \ \  (0 \ \ {\rm otherwise})
\label{eq_PSF}
\end{equation}
where $\theta$ is the angular separation to the center and $\theta_{max} = \Delta \theta_{pix}*\sqrt{2}$
with $\Delta \theta_{pix}$ being the pixel size (12 arcminutes for ROSAT and 3.7 arcminutes for {\small HEALPIX}).This expression results from fitting random orientations of pixels of size $\Delta \theta_{pix}$ and containing the central position of the source. This position can be anywhere (i.e is also a random variable) within the pixel.

The final average map obtained after stacking the areas $2.5^{\circ}\times 2.5^{\circ}$ around the 750 clusters is shown
in figure \ref{fig_RASS_2D}. The angular extent of the average cluster is very similar to the kernel representing the
RASS pixel plus stacking process which is not surprising given the big pixel size of the RASS pixel (12 arcminutes) making
most of the clusters appear as unresolved or barely resolved. 
Figure \ref{fig_RASS_1D} shows the one-dimensional profile of the stacked X-ray fields and a comparison with the effective 
PSF resulting from combining the intrinsic PSF of ROSAT and the two smoothing kernels due to the stacking+pixelization process. 


\begin{figure}  
   \epsfysize=18.0cm   
   \begin{minipage}{\epsfysize}\epsffile{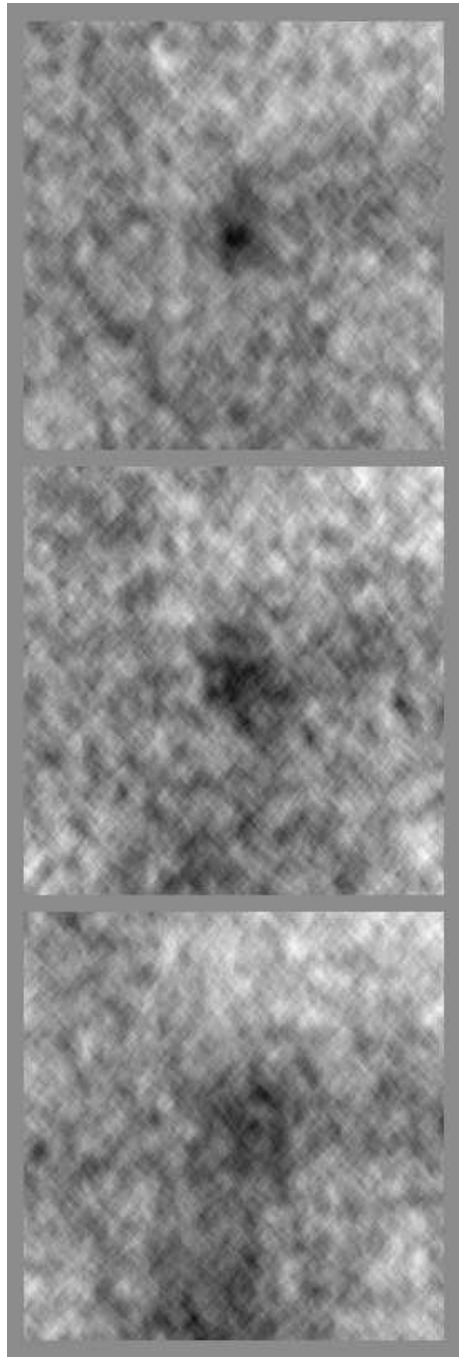}\end{minipage}  
   \caption{SZ in WMAP data. From top to bottom, $W$, $V$ and $Q$ bands. 
            Each panel shows the average signal obtained after stacking an area of 
            $2.51^{\circ}\times2.51^{\circ}$ around the clusters in the X-ray catalog. 
            Those clusters with bright radio sources have been masked. Note the very 
            clear decrement in the $W$ band. The decrement can still be seeing in the 
            $V$ and $Q$ bands but with less intensity due to the beam smearing.}  
   \label{fig_WMAP}  
\end{figure}

\section{WMAP data}
We use three of the WMAP frequency bands, $Q$ (at 41 GHz), $V$ (at 61 GHz) and $W$ (at 94 GHz) to search for the signal from
known galaxy clusters in the average stacked map. A more detailed description of the WMAP five year data and its main results 
can be found in Hinsaw et al. (2009). Due to beam dilution effects, we expect to have the 
strongest SZE signal in the $W$ band which has superior angular resolution. The average beam profile of WMAP 
is well approximated by a Gaussian in all three bands with $FWHM = 0.51^{\circ}$ for $Q$, $FWHM = 0.35^{\circ}$ for 
$V$, and $FWHM = 0.22^{\circ}$ for $W$. In addition to this beam, the pixelization+stacking procedure introduces an additional 
smearing function as described in section \ref{section_stack}. The combination of the intrinsic WMAP beam and the 
pixelization+stacking procedure produces a slightly different effective beam (see below). 
In order to remove bright radio sources, we use the kp0 mask.

\begin{figure*}  
   \begin{flushleft}
   \epsfysize=7.0cm   
   \epsfxsize=18.0cm   
   \begin{minipage}{\epsfysize}\epsffile{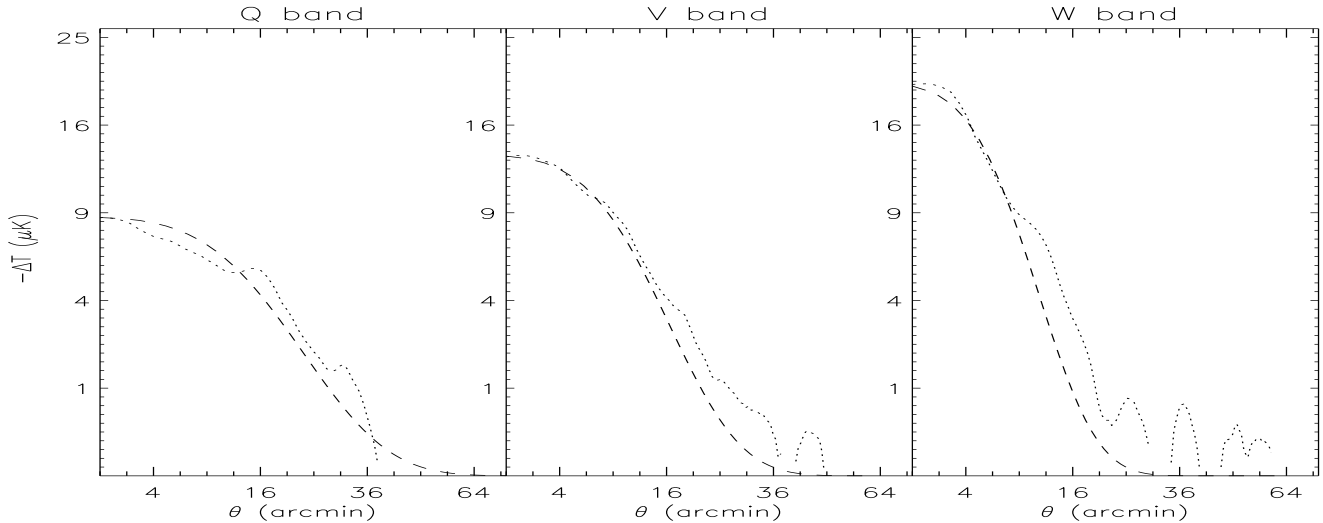}\end{minipage}  
   \caption{One dimensional average cluster profiles (dotted lines) compared with the effective beam profile (dashed lines).  
            The effective beam profile is a convolution of the intrinsic WMAP beam and the pixelization+stacking smearing function. 
            The effective beam has been normalized to the maximum amplitude of the average cluster profile. 
            The $x$ and $y$ axis grow as the square of the scale and temperature respectively 
            to better show the details in the central part and tails. For instance, 
            the maximum decrement is achieved in the $W$ band with a amplitude of $\approx 4.5^2 = 20 \ \mu K$ 
            (on the same plot  $16  \ \mu K = 4^2$).
            Note how the average cluster profile seems to be {\it resolved} in the $W$ and $V$ bands.}
   \label{fig_prof1}  
   \end{flushleft}
\end{figure*}

Following a process similar to the one described in the previous sections, we stack the 750 fields around the X-ray clusters and 
for each band ($Q$, $V$ and $W$). The result is shown in figure \ref{fig_WMAP}. A clear signal is seen in 
the three bands at the position of the clusters. As expected, the signal is stronger in the $W$ band, mostly due 
to its better angular resolution. Around the central position of the clusters we can see a degree scale structure in the three 
bands with hotter and colder fluctuations. This is most likely the result of stacking the CMB anisotropies in the 750 fields of 
view. We tested this hypothesis by stacking the same number of fields of view of a simulated map of the CMB. This simulation 
was done with the same angular resolution as the WMAP $W$ band. The resulting map shows 
similar patterns to the ones observed in figure \ref{fig_WMAP} and with the same strength. The average of the stacked simulated CMB was 
$4.8 \ \mu$K and the stacked map had a dispersion of $2.3  \ \mu$K, similar to the observed dispersion in the borders of the maps in 
figure \ref{fig_WMAP}. As an additional test we looked at the stacking of the WMAP fields with the 80 
largest (most positive fields or MPF) and 80 smallest 
(most negative fields or MNF) values in a central region of 7 arcminute  radius. The motivation for this exercise is to look for possible point source 
contamination in the stacked MPF. The same exercise was made with the simulation of the CMB described above 
(with no instrumental noise). In the case of the CMB simulation, the stacked MPF shows a large scale positive fluctuation with a scale of 1-2 degrees and 
an amplitude of about 100 $\mu$K. The stacked MNF shows a similar structure but with opposite sign.  
When looking at the WMAP data, the stacked MPF shows a structure similar to the one seen with the simulated CMB. 
No signs of point source contamination are observed in these fields indicating that if point source contamination occurs it must be sub-dominant with respect to the 
CMB. In the case of the stacked MNF a similar structure but with negative sign was observed. However, in this case a small scale negative fluctuation is seen at the 
center of the field of view, specially in the $W$ band indicating that among the sub-sample of 80 MNF, there is a significant contribution from the SZE. 
 

When looking at the one dimensional profiles (obtained after averaging in rings around the center), 
we can appreciate in  figure \ref{fig_prof1} how a comparison with the effective beam 
indicates that the average cluster profile is resolved in the $W$ band as well as in the $V$ band. As noted above, the dispersion in the wings of the 
one dimensional profile is consistent with the $2.3  \ \mu$K dispersion of the simulated (and stacked) CMB maps. The profile observed 
in the lower resolution $Q$ band is compatible with the profile (dashed line) expected for an unresolved source convolved with the combined PSF. 
Resolving the average cluster profile with WMAP 
data means that either the average signal is dominated by low redshift clusters or that there is significant gas in the outskirts of 
galaxy clusters (a third option is of course a combination of the two). The second option is interesting since if confirmed it could 
be an indication that a fraction of the missing baryons could be in these regions. 

Another interesting result is the fact that if we do not use the kp0 mask, and hence leave the bright microwave sources in the WMAP fields, the SZE signal 
disappears in the $Q$ band and is significantly reduced in the $V$ band while remaining basically unchanged in the $W$ band.
This suggests that significant radio source contamination from non-subtracted radio sources might still occur 
in the lower frequency bands while it should be much 
smaller in the $W$ band. This, however, does not rule other kinds of point source contamination (like infrared sources), as the point 
sources on which the kp0 mask is based are mostly (if not all) radio sources. 

We should note that the average of the stacked fields is not zero. Because we are using the uncleaned maps, there is still a 
non-negligible and positive contribution from the Galaxy. Before comparing the SZE profiles with models is necessary to subtract the 
Galactic contribution to the stacked fields. Since this value is unknown, we subtract a constant from each map such that the one dimensional 
profiles average to zero in the tails of the one dimensional profiles (see figure \ref{fig_prof1}).

\section{Model comparison}

\begin{figure}  
   \epsfysize=6.0cm   
   \begin{minipage}{\epsfysize}\epsffile{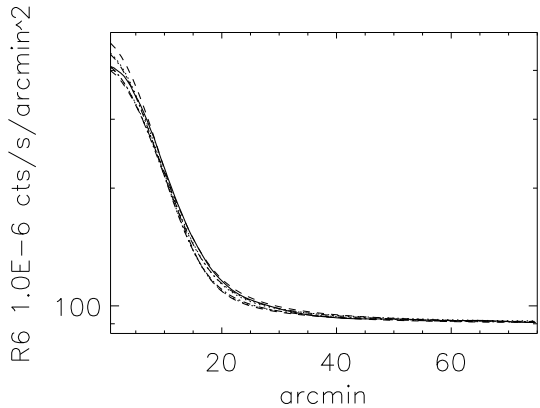}\end{minipage}  
   \caption{Predicted versus observed (solid line) one dimensional profile for RASS clusters. The thick lines represent models 
            that give a relatively good fit to the WMAP data while the thin lines are used for models who give a better fit to RASS 
            data. The models represented as thin lines correspond to (dotted line) the standard $\beta$-model (with $\beta=2/3$),  
           concentration parameter $p = r_v/r_c = 5$ and a core radius equal to two times the one predicted by the 
           self similar model.m This is denoted model $M_1$ in table 1. 
           The thin dashed line is for $M_2$ in the same table. The thin dot-dashed line is for $M_3$. 
           The thicker lines correspond to (dotted line) $M_4$,  
           thick dashed line $M_5$ and thick dot-dashed line $M_6$.}
   \label{fig_RASS_1D_2A}  
\end{figure} 

In this section we compare different models with the X-ray average emission from RASS maps and the SZE emission in WMAP data.
It is important to note that in constructing the RASS average profile, we have masked out the central part of the 314 
clusters identified in the Voges et al (1999) catalog. The masking does not affect the computation of the central density, since 
the X-ray fluxes were derived using the full resolution PSPC data of Snowden et al. (1997) and therefore do not mask the central regions of the maps. 
However, we should not expect a perfect match between the predicted and observed average signal at the cluster center due to very non-linear 
processes such as cooling flows that are not included in our model. In fact, due to these non-thermal effects, we should expect a good 
model to predict slightly lower X-ray flux than the observed one since we are assuming that all X-ray flux is due solely to thermal 
bremstrahlung (and emission lines). 
The wings of the average cluster profile should not be affected as much by these highly non-linear phenomena and hence 
they should be more useful to set reliable limits on the models. 
Any realistic model that we consider should predict correctly (or even better slightly underpredict) 
the average signal seen in the RASS maps at radii between $\sim 15 - 40$ arcmin. 

\begin{figure}  
   \epsfysize=6.0cm   
   \begin{minipage}{\epsfysize}\epsffile{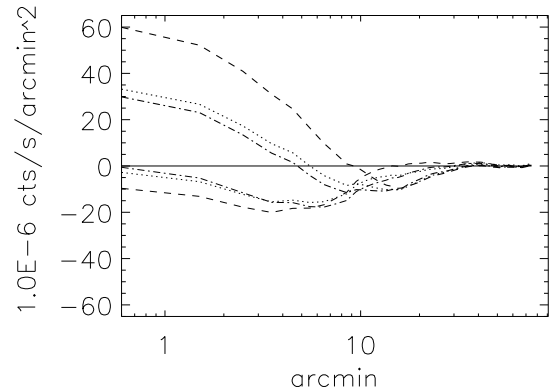}\end{minipage}  
   \caption{Same as in figure \ref{fig_RASS_1D_2A} but showing the difference of the predicted minus observed one dimensional profiles 
            for RASS clusters.}
   \label{fig_RASS_1D_2B}  
\end{figure}

\begin{table}    
\caption
{
All $\beta$-models have  $\beta=2/3$. 
SS stands for {\it Self-Similar} model and Alt for {\it Alternative}. 
The third column ($r_c/a$) shows the core radius (for the $\beta$-model) or the scale factor $a$ (for the AD08 model). 
SSx4 means the self similar normalization has been multiplied by a factor four and Altx1.5 means that the normalization 
of the Alternative model has been multiplied by a factor 1.5. SSx10 in the AD08 model means that the scale factor $a$ is 
equal to ten times the self-similar core radius model. 
}                                                         
\begin{tabular}{lcccc}                                                          
\hline                                                                          
Model      & Type    &   $r_c \ or \  a$   &  $p$   & Fits  \\      
\hline                                                                          
$M_1$      & $\beta$ &   SSx1.5  &   5    & X-ray \\                                    
$M_2$      & $\beta$ &   Altx1.5 &  10    & X-ray \\                                    
$M_3$      &  AD08   &   SSx10   &   5    & X-ray \\                                    
$M_4$      & $\beta$ &   SS      &   5    & WMAP \\                                    
$M_5$      & $\beta$ &   Alt     &   5    & WMAP \\                                    
$M_6$      &  AD08   &   SSx13   &  10    & WMAP \\                                    
$M_7$      &  AD08   &   SSx13   &   9    & Both ? \\                                    
\hline                                                                          
\end{tabular}          
\label{tabSignal}                                                               
\end{table}                                                                     
 
Modeling the average X-ray emission from clusters as seen by ROSAT PSPC detector involves a series of 
steps that need to be done carefully. 
First, we need a model for the gas distribution in three-dimensions (or 3D) that can be integrated to compute the X-ray luminosity 
and flux. The models considered in this work consist of the standard $\beta$-model and the polytropic model AD08 
as described earlier. For the X-ray emission we use a Raymond-Smith model (Raymond \& Smith 1977) 
with a metallicity of 0.3. This model 
includes emission lines which can change significantly the X-ray flux for lower temperature clusters (T of  few keV or less). 
The total X-ray flux is then converted into observed flux in the 0.1-2.4 keV band. This predicted flux can be 
compared with the observed one. This process involves two corrections, first the band correction that transforms bolometric 
flux into flux in a given band (in the rest frame of the cluster) and second the $K-$correction that accounts for the appropriate 
redshift distortions. For these corrections we used the values given in tables 3 and 5 in B\"ohringer et al. (2004). 
In order to compare with the average profile obtained from RASS maps, the flux needs to be converted into counts per second 
(or cps) in the R6 band. For this conversion we use the code XSPEC (Arnaud 1996) to fold the 
same Raymond-Smith model with the response matrix of the PSPC detector and in the appropriate 
range of energies of the RASS R6 band (0.9-1.3 keV). The conversion factor depends on the plasma temperature and cluster redshift. 
For temperatures above a few keV and low redshift this conversion factor is about 1 cps $= 2\times10^{-12} \ \ erg/s/cm^2$. 
We have assumed a negligible hydrogen column density (most clusters see less than $1.5 \times 10^{21} \ cm^{-2}$ $n_H$). 
The error introduced by neglecting the hydrogen absorption is small ($\sim 10\%$) for the range of 
temperatures, energies and column densities used in this work. 

Given a particular assumption for the gas distribution in a cluster at redshift $z$, a model for the core radius $r_c$ 
and the central electron density $n_o$, we can compute the expected X-ray flux and SZE by integrating equations 
(\ref{eq_DeltaT}) and (\ref{eq_Sx}). The integration is carried out up to a maximum radius $R_{max}$. The ratio between this maximum radius and the core radius is denoted by the parameter $p$ and both the X-ray and SZE signals de4pend on the value of $p$ although significantly less than on the choice for $r_c$.
The 3D models are projected along the line of sight and we obtain two-dimensional (or 2D) maps of the SZE and the X-ray emission. The integrated 
X-ray flux is compared with the observed X-ray flux for each cluster. Since the integral in equation (\ref{eq_Sx}) 
is proportional to $n_o^2$, our initial model can take any arbitrary value for $n_o$ and then be re-scaled 
by the appropriate factor that makes the predicted X-ray flux match the observed one. 
The root-square of this scaling factor is the multiplied by the SZE 
emissivity to produce the predicted SZE signal for that cluster. The 2D maps are then stacked together and an average profile can be obtained for the 
SZE and X-rays. For building the X-ray average profile we first convert the 2D flux maps into their corresponding 2D cps maps 
in the 0.9-1.3 keV band as described above. This conversion is done before stacking the maps since the conversion factor depends on 
the redshift and temperature of the clusters. As mentioned earlier, we remove the central 3.5 arcmin radius region in 
those clusters that were identified in the BSC catalog of Voges et al. (1999). 
After convolving the 2D maps with the appropriate effective PSF we can compare our 
predicted signal with the observed one in ROSAT and WMAP. The effective PSF for ROSAT was described in section \ref{section_stack}. 
For WMAP the effective PSF is again the combination of the intrinsic WMAP beam and the smearing introduced by the pixelization+stacking 
procedure. This smearing function can be described again by equation (\ref{eq_PSF}) but with $\Delta \theta_{pix} \approx 6.7$ arcmin 
which is the pixel size used for WMAP data (see equation \ref{eq_PSF} and description afterwards).

\begin{figure}  
   \epsfysize=6.0cm   
   \begin{minipage}{\epsfysize}\epsffile{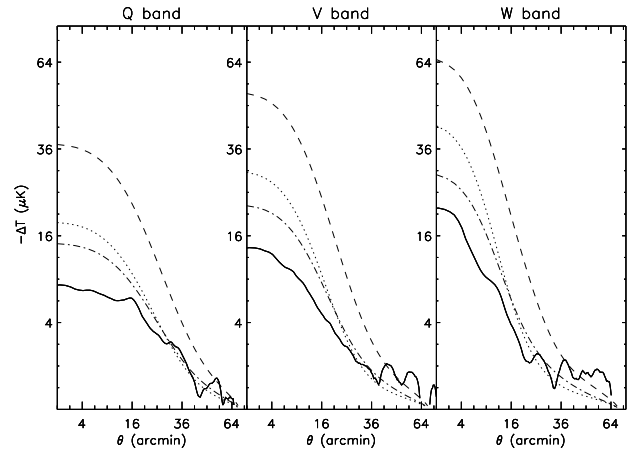}\end{minipage}  
   \caption{Models from figures \ref{fig_RASS_1D_2A} and \ref{fig_RASS_1D_2B} which fit the X-ray data well compared with WMAP data. 
            Note how the $\beta$-models (dotted and dashed curves) over-predict significantly the observed WMAP average decrements. 
            The models correspond to $M_1$ (dotted), $M_2$ (dashed), and $M_3$ (dot-dashed) from table 1. The vertical axis shows the 
            negative SZE decrement with opposite sign.}
   \label{fig_WMAP_1D_A}  
\end{figure} 

\begin{figure}  
   \epsfysize=6.0cm   
   \begin{minipage}{\epsfysize}\epsffile{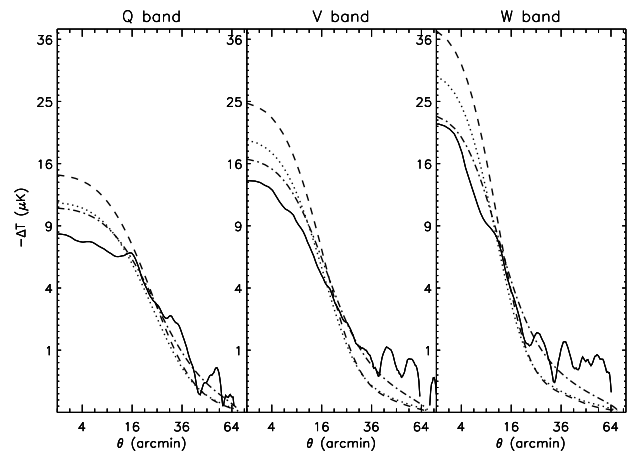}\end{minipage}  
   \caption{SZE fitting models from figures \ref{fig_RASS_1D_2A} and \ref{fig_RASS_1D_2B} compared with WMAP data. 
            The models correspond to $M_4$ (dotted), $M_5$ (dashed), and $M_6$ (dot-dashed) from table 1.}
   \label{fig_WMAP_1D_B}  
\end{figure} 

In figure \ref{fig_RASS_1D_2A} we show different models that seem to agree well with the average X-ray profile. These models are 
listed in table 1. We have chosen two sets of models. The first set (represented with thin lines in figure \ref{fig_RASS_1D_2A}) 
reproduce particularly well the average X-ray profile observed in RASS maps. The set with thicker lines still reproduce more or less this 
profile but seems to deviate in the wings, predicting less X-rays than observed. As we will see later, 
these models (thick lines) will fit the SZE WMAP profiles better than the thin line models. For a better comparison, the same models are represented 
again in figure \ref{fig_RASS_1D_2B} but this time showing the difference between the observed profile and the predicted ones. 
When the same models are compared with the SZE we find that it is nearly impossible to find a model that can fit both data sets 
simultaneously (see figures \ref{fig_WMAP_1D_A}  and \ref{fig_WMAP_1D_B}). 
Models $M_1$, $M_2$ and $M_3$ give a good fit to the X-ray data but a bad fit to the SZE. Conversely, models $M_4$, $M_5$ and $M_6$ give a 
better fit to the SZE data but a poorer fit to the X-ray data. We observed that large core radii are needed to fit the X-ray data. As described 
earlier, the SZE grows as $r_c^{3/2}$ when the gas profile is constrained with the X-ray fluxes. Consequently, the large $r_c$ needed to fit 
the X-rays predict a large amount of SZE. Smaller values for $r_c$ push the SZE down to levels compatible with the SZE seen in WMAP but this 
smaller $r_c$ would fail in fitting the wings of the X-ray profile. It seems that some level of compromise will be needed to reconcile both data sets. 
Given the assumptions made in the modeling of the X-ray average signal, it is reasonable to assume that we do not expect a perfect match between 
the observed X-ray average profile and the predicted one in our model. For instance, no cooling flows or shock fronts were included in our model. 
In table 1 we list one model ($M_7$) that makes a relatively good fit to both data sets (see figure \ref{fig_Best_PS} below) 
but is not the best fitting model to either of the 
two data sets individually. The model $M_7$ under-predicts slightly the X-ray emission over most of the radial profile. 
This, however, is a desirable feature since we should not expect a perfect fit for a model that does not include 
some phenomena like cooling flows, shock waves in the gas, or cold fronts. These mechanisms boost the X-ray emission without requiring more 
gas. Instead. a good model should slightly under-predict the X-ray emission in order to leave some room for these highly non-linear processes 
to fill the gap. The same processes are expected to have little impact on the SZE. 
In the next section we will see that this model has another very interesting feature. 
It is reasonably consistent with estimates of the expected level of contamination from point sources in galaxy clusters. 
This model can be considered as a concordance model in the sense that it delivers a reasonable fit consistent with the 
available observations.   Note that no $\beta$-model does as well.

\section{Possible contamination from radio and/or infrared sources}\label{section_PS}  
\begin{figure*}  
   \begin{flushleft}
   \epsfysize=4.6cm   
   \epsfxsize=17.2cm   
   \begin{minipage}{\epsfysize}\epsffile{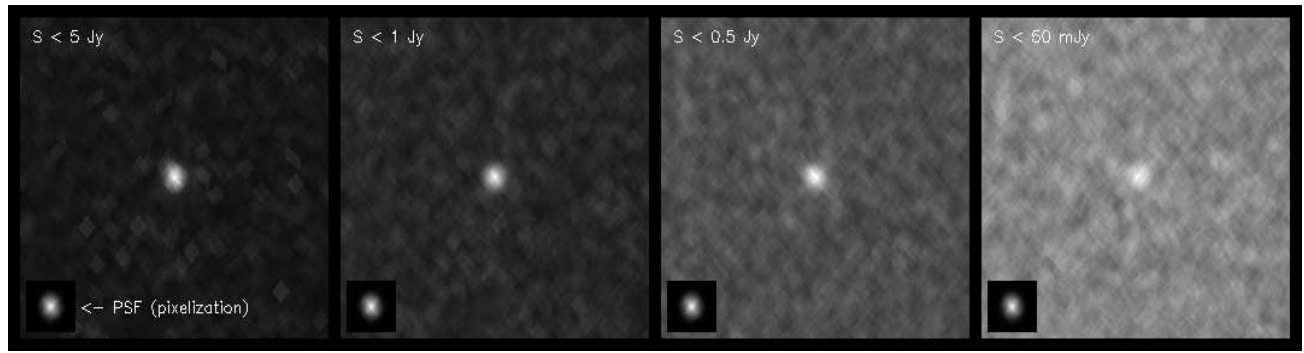}\end{minipage}  
   \caption{Stacked NVSS sources at the positions of the clusters. From left to right, sources with fluxes smaller than 5 Jy, 
            fluxes, smaller than 1 Jy, 0.5 Jy and 50 Mjy respectively. Also shown is the smearing introduced by the stacking+pixelization.}
   \label{fig_NVSS}  
   \end{flushleft}
\end{figure*} 

Previous papers have reported that the amount of SZ observed in WMAP data is less than expected 
Atrio-Barandela et al. (2008), Lieu et al. (2006), Afshordi et al. (2007), Bielby \& Shanks (2007). 
We confirm this result by looking at the temperature fluctuations in clusters. 
Also, the lack of signal seems to be more prominent in the outskirts of the cluster suggesting that the pressure profile falls 
faster than predicted by the $\beta$-model (Atrio-Barandela et al. 2008). A possible mechanism to explain both observations 
would be a spatially uniform distribution of radio galaxies inside galaxy clusters. Such a uniform distribution would make the integrated 
radio signal more relevant at larger radii and also the overall positive contribution would reduce the total integrated SZ signal. 
We check this hypothesis using a catalog of radio sources (see below). 
It is known that some level of radio emission exist in galaxy clusters at the frequencies of WMAP (see discussion of the results of  Lin et al. 2009 
below). In an earlier work, Cooray et al (1998) made a survey of radio galaxies in clusters at 28.5 GHz. 
They found that there were 4 to 7 times more sources than predicted from a low-frequency survey in areas outside galaxy clusters. 
They concluded that contamination from radio sources could be a serious threat to mm or cm SZE surveys.  In a later work, 
Coble et al (2007) confirmed these results (also at 28.5 GHz) with a larger sample of sources in galaxy clusters. 
Lin et al. (2009) made an even more extensive study including also higher frequency bands. One of their most relevant conclusions 
for this work is that a significant fraction of radio sources show inverted or flat spectrum at frequencies beyond 10 GHz. 
The results of Lin et al. (2009) will be discussed in more detail later. 
White \& Majumdar (2004) predicted the average level of contamination of radio and infrared sources at mm wavelengths 
based on extrapolations from existing data.  In particular, they predict that in 
the $W$ band the confusion noise from infrared sources should equal the confusion noise from radio sources (with both equal to 
approximately 10 $mu$K for a one arcminute beam). Other works predict the contribution of point sources based on analytical models 
(Toffolatti et al. 1998, De Zotti et al. 2005). 
These works do not focus on galaxy clusters but rather on the average contribution of point sources to the CMB. Nevertheless 
they are still useful to set a minimum limit to the contribution of point sources to galaxy clusters  
(though we expect an enhanced signal due to clustering). 
  
At WMAP frequencies, we should expect the calculated SZE  signal to over-estimate the observed SZE by some amount, the difference 
being due to point source contamination. In this section we try to estimate this level of contamination and combine it 
with the concordance model described above ($M_7$). 

\subsection{NVSS data}
The NRAO-VLA Sky Survey (or NVSS, Condon et al. 1998) catalog contains about 1.8 million sources 
with measured fluxes at 1.4 GHz. The NVSS survey has an effective resolution of
45 arcseconds, much smaller than the angular resolutions relevant for this work and its effect can be safely neglected for our
purposes. Using this catalog we build an all sky map using again the {\small HEALPIX} pixelization and the same pixel size as in
the WMAP data (6.8 arcmin pixel). The stacking process at the positions of the 750 clusters on our list is similar to the cases discussed above. 

In figure \ref{fig_NVSS} we show the result of the stacking procedure. NVSS bright sources can be seen at the center of 
the clusters in our catalog. We have divided the NVSS catalog into different flux bins and found that the NVSS 
sources seem to be strongly correlated with clusters down to fluxes of a few tens of mJy at 1.4 GHz. How much of this flux still 
remains at WMAP frequencies is still an open question but works like the ones presented in Lin et al. (2009) are helping to throw some 
light on this issue. From our results, however, we see no evidence of a spatially uniform distribution of radio sources inside the clusters. Instead, the 
radio sources seem to be concentrated in the center of the clusters (as found by Lin et al. (2009)). 

\subsection{Infrared data}
On the other side of the spectrum, we repeated the same exercise  but on the infrared side using the IRAS galaxy catalog from 
Rowan-Robinson et al. (1991). 
This catalog contains 17711 sources with measured fluxes at 12 $\mu$m,  25 $\mu$m, 60 $\mu$m,  and 100 $\mu$m. For the 
stacking we used the 100 $\mu$m fluxes as the corresponding frequency is closest to the WMAP ones. 
We found no significant flux overdensity in the stacked position of our clusters. 
This is not surprising given the small redshifts ($z < 0.02$) of the galaxies in the IRAS catalog. Also, the large difference 
between the IRAS largest wavelength (100 $\mu$m) and the WMAP ones makes it difficult to extrapolate the fluxes down to WMAP 
frequencies. A related and interesting work can be found in Zemcov et al (2007). Using SCUBA archival data around galaxy clusters, 
the authors try to recover the SZE signal at 850 $\mu$m. Their sample contains 44 clusters. Among their conclusions, the most revealing 
are the fact that a large fraction of their sample shows significant emission from active galactic nuclei (or AGN) at the center of the 
clusters and that for the rest, the derived amplitudes of the SZE increment are consistently higher than the amplitudes inferred from 
lower-frequency measurements of the SZ decrement. The disagreement could be easily explained if additional (i.e not accounted for) 
emission from the infrared galaxies in the cluster is still contributing substantially to the total flux. In Zemcov et al. (2007), the authors provided also a very interesting list of 83 candidate sources 
detected in galaxy clusters together with their fluxes at 850 $\mu$m. Fluxes in this 
sample range from 3 mJy to 70 mJy and with an average flux of 8.7 mJy. Assuming a $\nu^2$ law for the flux (see for instance Greeve et al 2008) 
to extrapolate to WMAP frequencies, this average flux translates into a flux smaller than 1 mJy in the $W$ band. We thus reach a different conclusion about the relative importance of radio and infrared sources for White \& Majumdar (2004). This disagreemet is due to the fact that Lin et al. (2009) focus on galaxy clusters and also their extrapolation is more accurate since their frequencies are closer to the $W$ band.

\subsection{Contamination from lensed background sources}
An interesting mechanism capable of increasing the flux inside a galaxy cluster with respect to the mean flux is lensing. 
Lensing of background galaxies by foreground clusters can create magnifications of a factor 10 or more. Big 
magnifications occur far more often around galaxy clusters than in non-cluster regions.  Also, in the mm to sub-mm 
band an interesting phenomenon occurs, the negative k-correction. A typical starburst galaxy will be 
seen with almost the same intensity at redshifts 2 and 5. This means that if one is able to reach a sensitivity 
enough to detect redshift 2 galaxies at these frequencies, the same sensitivity will allow one to see all these galaxies at even 
higher redshifts. Lensing of mm and sub-mm sources has been discussed in the past (a detailed analysis can be found in 
for instance Perrotta et al. 2003, see also Paciga et al. 2009) 
but from a statistical point of view and focusing on the number counts. 

\begin{figure}  
   \epsfysize=6.0cm   
   \begin{minipage}{\epsfysize}\epsffile{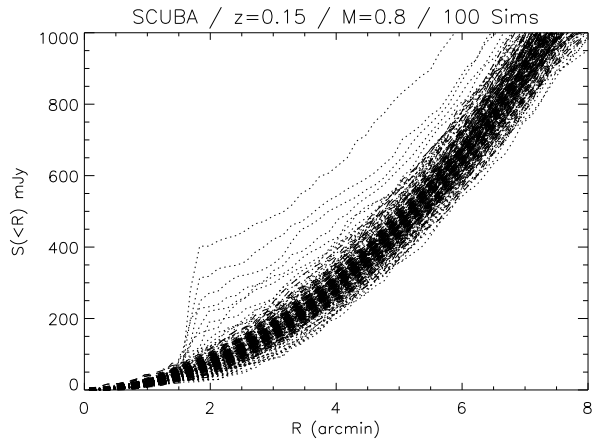}\end{minipage}  
   \caption{Predicted contamination from lensing for an illustrative model. The curves show the integrated flux from background sources 
            in a disc of radius $R$ centered in the lensing cluster. For a uniform distribution of background sources 
            the flux grows as $R^2$. Dashed lines show the integrated flux for the unlensed field of view. Dotted lines 
            are the corresponding flux for the lensed field of view. Each line corresponds to a different realization. 
            The cluster is located at redshift 0.15 and it has a mass of $8\times 10^{14} M_{\odot}$. The background 
            sources were assumed to be at redshift 2. A typical SZE flux is of the order of a few hundred mJy at the SCUBA wavelength (850 $\mu$m).}
   \label{fig_SCUBA}  
\end{figure}

In this section we focus on the possible contamination by lensed galaxies of the SZE. 
We use a toy model to quantify the level of contamination introduced by lensed galaxies. In our model we consider 
a cluster at a redshift of 0.15 (representative of the average redshift of our sample). 
Low redshift clusters have larger lensing cross sections. This means that the 
lensing effect  will be smaller for higher redshift clusters. The mass of the cluster is $8\times 10^{14} M_{\odot}$ 
and is simulated by superposing three NFW profiles with different centers. 
The superposition of the NFW halos adds some substructure to the model cluster but to first order 
it has spherical symmetry. In this case, lower masses will produce smaller magnifications. 
For this cluster we compute the magnification map assuming a redshift of the source of 
$z=2$. Observations of mm and sub-mm sources with SCUBA at 850 microns (Clements et al. 2008) and MAMBO at 1.2 mm 
(Bertoldi et al. 2007) show that the peak of the distribution is 
around $z=2$. Given the redshift of our toy cluster, the magnification map depends weakly on the redshift of the sources. 
We use the number counts derived with SCUBA (Coppin et al. 2006) and simulate a background of sources at redshift 2. 
The sources are distributed randomly and uniformly. We make 100 different realizations and for each one we obtain their lensed counterpart. 
Finally, we compare the lensed with the unlensed images. We take discs of increasing radius centered on the cluster and compute the 
integrated flux for each disc in the two images (lensed and unlensed). The result is shown in figure \ref{fig_SCUBA}. The unlensed 
curves correspond to the dashed lines and they grow with the square of the disc radius (as expected). The lensed integrated 
flux (dotted lines) behaves almost the same with the exception of a few realizations ($\approx 7\%$) where significant magnification occurs. 
The maximum magnification occurs around the critical curves. Also, we found that a number of realizations 
($\approx 10\%$) suffer the contrary effect (demagnification of the integrated flux) caused by a large magnification of an 
under-dense (or empty) region. 
With this toy model we can conclude that in the case of SCUBA sources, the lensing effect can increase significantly 
the flux in the cluster compared with the SZE flux in only a few percent of the cases. Also, by taking the average of 
a sample of clusters, magnification and demagnification effects compensate each other (at least partially) making the 
average lensing contribution even smaller. A study combining SCUBA (850 $\mu$m) and MAMBO (1200 $\mu$m) 
sources revealed that the typical spectral index for 
these sources is $\alpha=2$ (Greve et al. 2008) where $S \propto \nu^{\alpha}$. 
This implies that at lower frequencies the lensing effect is 
expected to be even less relevant as there should be a smaller density of bright background sources (and hence, it would be 
more unlikely to have a source magnified). For example, from the number counts given in Bertoldi et al (2007), 
the density of sources at 1.2 mm is about an order of magnitude less than the density of SCUBA sources at 1 mJy.  
For experiments operating at even lower frequencies (like the Atacama Cosmology Telescope, or ACT, Kosowsky 2006) the 
average impact of lensed sources should be even smaller with only a few percent of clusters showing a significant 
contamination to their SZE flux. The same conclusion applies to WMAP frequencies where 
the contamination from lensing effect should be negligible. 

\section{SZE plus point source model}\label{section_PS_II}  
In this section we try to fit the SZE profile considering a model that includes a point source (or PS) 
at the center of the cluster. Since the PS contributes as a positive $\Delta T$ and the SZE as a negative one, a partial cancellation 
of the later signal might occur. 

\begin{figure}  
   \epsfysize=6.0cm   
   \begin{minipage}{\epsfysize}\epsffile{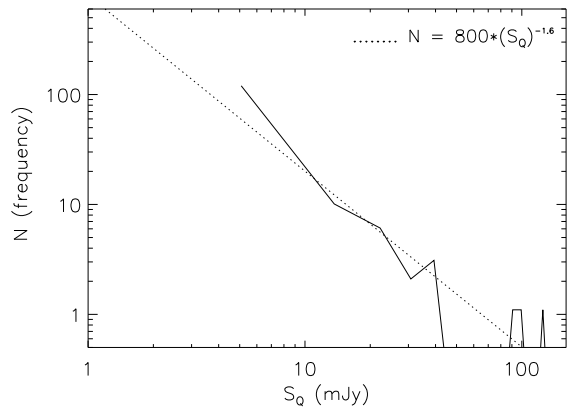}\end{minipage}  
   \caption{Radio sources from Lin et al. (2009). The solid line shows a histogram of the 
            radio sources observed in galaxy clusters by Lin et al (2009). The mean flux of these sources is 9.2 mJy 
            in the Q band. The dotted line represents a model for the number counts which extrapolates the fluxes down to 1 
            mJy. The mean flux for this model is 10.4 mJy. Lower minimum boundaries for the flux will result in a 
            lower average flux.}
   \label{fig_Lin}  
\end{figure}

Lin et al. (2009) observed a sample of 139 radio selected sources inside galaxy clusters. These sources were observed 
using the Very Large Array (or VLA) at frequencies between 4.9 and 43 GHz (note that the highest frequency matches the 
frequency band of the $Q$ channel in WMAP). They found that most radio sources in clusters lay near the center of their 
host cluster and that  a significant amount of the sources in clusters show a flat or 
even inverted spectrum that would increase their flux at higher frequencies compared with the predicted flux using standard 
frequency extrapolations. We compute the average flux of their sample at the highest frequency (43 GHz) and find 
this to be 9.2 mJy, or 12.4 mJy if we consider only the flatter spectrum
unresolved and core sources. 
Some of the fluxes at 43 GHz given in Lin et al. (2009) were upper limits on the flux. 
Also, the minimum flux in their sample is a few mJy. We made a fit to their sample (see figure \ref{fig_Lin}) using a simple 
power law and recomputed the average flux extrapolating down to 1 mJy. The average flux for the model was 10.4 mJy. These numbers give 
us an idea that we should expect on average a PS contribution of about 10 mJy per cluster in the $Q$ band. This valuable information can be used 
to set an additional constrain in our model. 
Also in Lin et al (2009), the authors extrapolated the fluxes observed at lower frequencies to 90 GHz. Two extrapolations were done using 
different assumptions for the spectral index. The average flux of the sources in their table 3 at 90 GHz is 4.4 and 6.2 mJy for the 
two extrapolations. We should keep in mind however, that this extrapolation is highly uncertain and should be used with caution. 
Nevertheless, it is useful to establish that we can expect contributions of the order of a few mJy from radio sources at 
frequencies of 90 GHz (roughly, the $W$ band in WMAP). 

Regarding infrared sources, the situation is less clear as there are very few observations of infrared sources in galaxy 
clusters. As noted earlier, from the results of Zemcov et al. (2007) using SCUBA data, 
one can infer a contribution of less than 1 mJy in galaxy clusters from thermal dust emission. 
Other mechanisms (or the presence of colder gas in galaxies) could boost this flux to higher limits but this 
remains unclear so far. 

An interesting result can be found in Granett et al. (2008). In this paper the authors stack CMB regions around 
50 supervoids and 50 superclusters found in the Sloan Digital Sky Survey (or SDSS, Adelman-McCarthy et al. 2008). 
They found a colder average signal around the voids and a hotter average signal round the superclusters. 
Their interpretation of the results is that the colder 
and warmer signals are due to the integrated Sachs-Wolfe effect (Sachs \& Wolfe 1967). 
This is certainly one of the possible explanations but one could argue 
as well that the colder region is explained by a lack of point sources in the supervoid region and the warmer region by an excess of 
point sources in those regions. If the emission from point sources in these regions is dominated by mm sources with spectrum $\nu^2$ (Greve et al. 2008), their frequency 
dependence would be virtually indistinguishable from the ISW effect. Also, in Granett et al. (2008), the signal from superclusters should also be sensitive to the negative 
contribution from the SZE. Instead, a positive signal is observed. Whether this positive signal is due to the ISW or point sources in the 
superclusters is an issue that needs to be resolved in the future for instance by observing the same regions at higher frequencies. 

\section{Discussion}\label{section_discussion}
A model with a steeper profile (like the AD08 
considered here) is capable of simultaneously reproducing the X-ray brightness and SZE profiles 
better $\beta$-model. 
Lieu et al. (2006) predicted a large SZE signal in their sample of clusters. This is possibly due to 
the fact that they used an isothermal $\beta$-model to fit the X-ray emission. 
We can argue that part of the discrepancies between the expected SZE signal considered in the past and the apparently smaller signal seen by 
WMAP can be explained by considering models for the gas distribution that are steeper than the $\beta$-model. This conclusion 
is in agreement with recent findings based on high resolution X-ray data in galaxy clusters that seem to suggest a steeper profile 
than previously thought (Neumann 2005, Vikhlinin et al. 2005, 2006, Croston et al. 2008). 
Atrio-Barandela et al. (2008) also found evidence of a steepening profile using WMAP data and a cluster sample similar to ours but using 
a different analysis. In their paper they concluded that the gas profile was more consistent with a NFW-like profile than a $\beta$-model 
profile. Mroczkowski et al. (2009) has recently completed a study combining X-ray CHANDRA observations with SZE high resolution 
interferometric measurements obtained with the Sunyaev-Zel'dovich Array (or SZA). In their analysis, they used a model with steeper profile 
than the $\beta$-model profile for the pressure and they also found good agreement between the X-ray and SZE observations. The long baselines 
of the SZA also permit the detection of radio galaxies inside the clusters and thus an estimate of their contamination level. 
Mroczkowski et al. (2009) found that in their small sample of clusters, the radio source contamination was at the level of a few mJy 
at 31 GHz. These kind of studies, were high resolution X-ray and SZE are combined in the same analysis, will be able to determine 
a more accurate model for the gas distribution. The interferometric data used in Mroczkowski et al. (2009) does not allow 
a study of  the gas on scales larger than that set by the shortest baselines of the SZA. Future data from the Atacama Cosmology Telescope (ACT, Kosowski 2004) 
or the South pole Telescope (SPT, Ruhl et al. 2004) will complement the interferometric studies and extend the angular scale range. 
Also, the Planck satellite will provide a list of SZE measurements in many hundreds of clusters.  Most of the clusters used in the present
work (if not all) will be observed by Planck. Even though Planck will not be able to resolve most clusters, it will be useful to determine 
the total flux of each cluster which can then be directly compared with the X-ray flux. This kind of comparison will allow a better 
determination of the gas contents in clusters.

\begin{figure}  
   \epsfysize=6.0cm   
   \begin{minipage}{\epsfysize}\epsffile{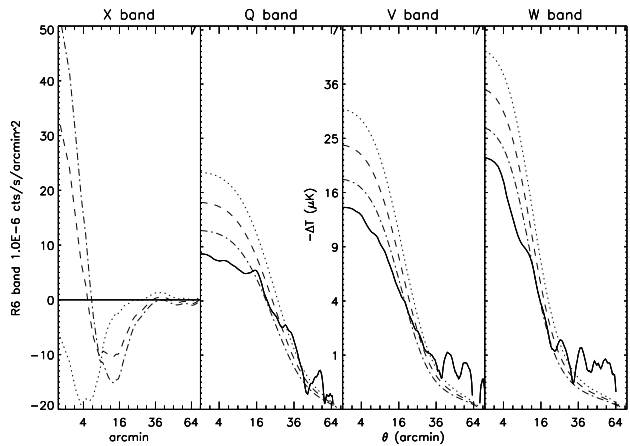}\end{minipage}  
   \caption{One dimensional profiles (solid lines) compared with three $\beta$-models.  
            From left to right, difference between models and X-ray RASS profile, 
            $Q$, $V$ and $W$ WMAP profiles. 
            The dashed line represents a $\beta=2/3$ model similar to $M_1$ but with $p=10$. The dotted line 
            is like the previous model but with $\beta=0.5$ and the dot-dashed line is for the 
            same model but with $\beta=0.8$}
   \label{fig_Betas}
\end{figure}

Perhaps one of the most revealing papers in the literature about the incompatibility of the $\beta$-model and 
X-ray+SZE data can be found in Hallman et al. (2007). In this paper the authors use simulated data of X-ray and SZE 
observations and conclude that {\it ``there is an inconsistency between X-ray and SZE fitted model parameters that leads to a bias 
in deduced values of the Compton parameter and gas mass when using isothermal $\beta$-models - X-ray and SZE radial profiles from the 
same galaxy cluster can not be fitted by identical isothermal $\beta$-models''.} 
Using our models we also find that it is hard to fit the X-ray and SZE profiles simultaneously using 
$\beta$-models. From the previous discussion it seems reasonable to assume that higher values for $\beta$ 
might agree better with the SZE measurements. We find that this is true but at the expense of failing to 
predict the X-ray measurements well. In figure \ref{fig_Betas} we show an example where we compare the 
average profiles measured by ROSAT and WMAP with models where we change the exponent $\beta$. Higher values 
of $\beta$ seem to fit better the WMAP SZE profiles (when the models are subject to the constraint that 
the total X-ray flux matches the measured X-ray flux). These models however, seem to under-predict 
significantly the X-ray emission in the wings of the profile. A similar behavior is observed when other 
parameters in the $\beta$-model, like the core radius or the parameter $p$ are varied. Reducing the 
core radius (or $p$) brings the SZE down but at the expense of failing to reproduce the wings 
in the X-ray profile. Although the $\beta$-models do not seem to agree well with both observations (X-ray ad SZE) we find that in general,  $\beta$-models with large values for $\beta$ perform better (than models with smaller $\beta$) when compared to both data sets simultaneously. A model like the AD08 which is intrinsically steeper (or other similar ones) is able to reproduce even better the X-ray and SZE profiles simultaneously. 

\begin{figure}  
   \epsfysize=6.0cm   
   \begin{minipage}{\epsfysize}\epsffile{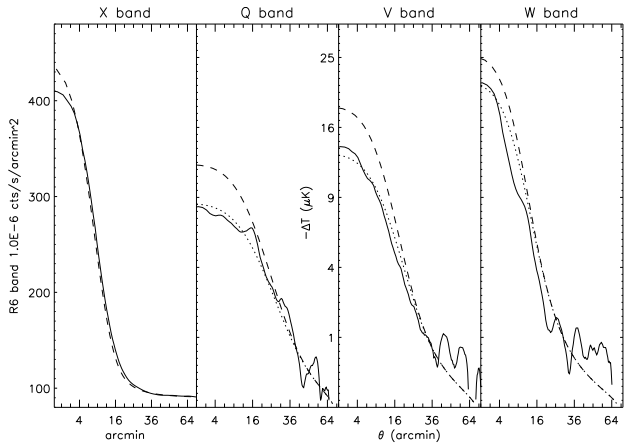}\end{minipage}  
   \caption{One dimensional profiles (solid lines) compared with the AD08 $M_7$ model (dashed line). 
            From left to right, X-ray RASS profile, $Q$, $V$ and $W$ WMAP profiles.
            The dotted lines in the $Q$, $V$ and $W$ bands show the AD08 $M_7$ {\it concordance} model minus the contribution 
            of a point source at the center of the cluster 
            with fluxes of 16, 26 and 18 mJy in $Q$, $V$ and $W$ bands respectively.}   
   \label{fig_Best_PS}
\end{figure} 

When adopting our concordance model ($M_7$ in table 1), we still found evidence for missing SZE in WMAP data. 
We attempt to explain this with a model for the contamination from 
point sources (radio and infrared) inside galaxy clusters. We estimated that infra-red sources should give a very small contribution 
in flux of the order of 1 mJy or less at WMAP frequencies. Radio sources, on the other hand could contribute with several mJy in all 
WMAP bands although this estimation is generally  derived from extrapolations made from lower frequency observations. In the $Q$ band a direct measurement of the average flux from point 
sources in a sample of clusters was made by Lin et al (2009). Radio sources are expected to contribute about 10 mJy on average in this band. 
A fit combining the AD08 model and a point source contamination of 16 mJy seems to render a god fit to the data in the $Q$ band (see figure \ref{fig_Best_PS}). At higher 
frequencies it is still not clear whether the difference between the model and the data is caused by point sources, a different gas profile 
or other unidentified mechanism. Given the actual constraints on point source contamination (about 10 mJy in the $Q$ band and a several 
mJy in the $W$ band) it may not be possible to reconcile the galaxy cluster X-ray and SZE observations without 
invoking more exotic or elaborated models for the gas distribution in galaxy clusters. 
Some of our models were able to fit well the WMAP SZE profiles and 
simultaneously be consistent with the expected level of contamination from point sources in $Q$, $V$ and $W$ bands. 
These models though, seem to under-predict the X-ray emission from clusters at intermediate radii. The models that perform better 
at these radii seem to over-predict the SZE signal in WMAP. In this case, estimates of point source contamination provide some help to reconcile the model with the observations. In figure \ref{fig_Best_PS} we show one of those models where the point source contamination is 16, 26 and 18 mJy in $Q$, $V$ and $W$ bands respectively. A $\beta$-model with a good fit to the X.-ray profile would require point source contamination several times higher than these values which would contradict the measurements of Lin et al. (2009). Another interesting conclusion that can be derived from figure \ref{fig_Best_PS} is the fact that the point source contamination seems to be largest in the $V$ band, in contradiction with previous expectations that extrapolate the radio emision to larger frequencies and the infrared emission to smaller ones. This might be an indication that the cluster model in figure \ref{fig_Best_PS} still needs to be improved but it also leaves the door open to future surprises in the frequency range 60-70 GHz. Data from Planck will help to answer this question in the incoming years. 

There are other possibilities not explored in this paper to explain a possible lack of SZE signal. 
One is the local clumpiness of the gas around smaller dark matter sub-halos in the galaxy 
cluster. This clumpiness would increase the amount of X-rays produced with the same amount of gas. Hence, a fit to the X-ray 
surface brightness with a smooth model (like the $\beta$-model or AD08 models) will again over-predict the amount of gas responsible 
for the X-ray emission and therefore the SZE produced by the same gas. A similar over-prediction will occur when shocks and other 
non-thermal phenomena occur in galaxy clusters that boost the X-ray brightness surface. 
CHANDRA and XMM observations reveal that some clusters show these features 
(Markevitch et al. 2000, McNamara et al. 2001, Fabian et al. 2003, Gioia et al. 2004, Ma et al 2009).  
Lin et al (2004) discuss also another interesting scenario in which preheating of the gas in clusters 
occurs at high redshift ($z \approx 2$). This mechanism was studied earlier by Kaiser (1991), and Evrard \& Henry (1991).
Such heating reduces the SZE flux by a factor of 2-3 on scales of a few arcminutes.

\begin{figure}  
   \epsfysize=6.0cm   
   \begin{minipage}{\epsfysize}\epsffile{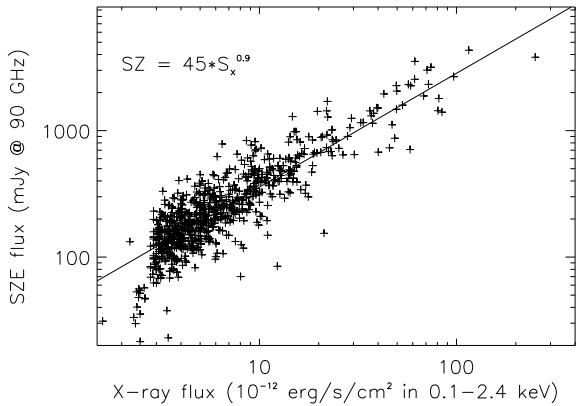}\end{minipage}  
   \caption{Predicted SZE flux vs measured X-ray flux. The predicted SZE was computed using the concordance model ($M_7$). 
            The solid line represents a scaling law (see text). }

   \label{fig_SZ_Sx}
\end{figure}

One other interesting conclusion can be derived from our results. Earlier works have predicted and/or measured 
the existence of tight scaling relations 
between X-ray and SZE derived quantities (McCarthy et al. 2003, Morandi et al. 2007, Bonamente et al. 2008, Yuan et al. 2009). 
The use of scaling relations has proved to be useful in the past to derive 
other quantities such as the cluster temperature. They are a powerful tool to investigate the physical properties of the 
clusters and their evolution in redshift. Using our concordance model as the best guess for the underlying gas distribution, we can 
predict the total SZE flux from each cluster and compare it with the observed X-ray flux. To first order, we should expect a clear correlation 
between the two since they both depend on the square of the distance. The X-ray depends on the luminosity distance and the SZE on the angular 
diameter distance and for low redshifts both are very similar. In figure \ref{fig_SZ_Sx} we show the correlation for the $M_7$ model. As expected, 
a clear correlation exists between the two quantities. The solid line represents the following scaling law;
\begin{equation}
Flux = 45 \times S_x^{0.9}
\end{equation}
where $Flux$ is the SZE flux in mJy at 90 GHz and $S_x$ is the X-ray flux in units of $10^{-12}$ erg/s/cm$^2$ and in the 
energy band (0.1-2.4) keV. The scatter around this scaling law is probably smaller than the real scatter in clusters but the scaling law 
itself can be considered representative as it was derived from the average properties of the cluster sample. 

Our results have profound implications for ongoing and future SZE surveys. Most of the predictions made in the past regarding 
the expected number of SZE detections for different experiments were made under the assumption that the gas profile can be 
well described by a $\beta$-model. Our findings suggest that this assumption is not valid and that a steeper gas profile would 
be more adequate. A steeper profile subject to the constraint that it has to be compatible with X-ray observations results in a smaller amount of SZE. Future data like ACT, SPT or planck 
will answer this and other interesting questions. 

\section{Conclusions}
By stacking WMAP 5 year data we are able to  see the average SZE signal from a large sample of galaxy clusters. These clusters 
are X-ray selected. A comparison of models with the average X-ray profile and the average SZE profile allows us to narrow down 
the range of possible models. Standard $\beta$-models are harder to reconcile with both data sets. A steeper profile, like the 
one suggested by Ascasibar \& Diego (2008) render a better fit to the data. None of the models considered is able to perfectly 
fit both data sets. Some error can be allowed in the fits, in particular in the X-ray profile, where non-thermal mechanisms or clumpiness in the gas can boost the 
X-ray emissivity without requiring more gas, we can find a {\it concordance} model that predicts a signal close to the observed ones (in 
X-rays and SZE) and also leaves room for point source contamination at roughly the level expected from radio and infrared extrapolations. 
We conclude that the X-ray and SZE measurements are consistent once a careful modeling of both data sets and a realistic level of 
point source contamination are taken into account. Based on the predictions made by the {\it concordance} model, we compare the SZE and X-ray 
fluxes and find a tight correlation between them. These kinds of correlations have been predicted in the past but never observed with a large 
sample of clusters. Future SZE data like Planck, and also high resolution experiments like ACT or SPT, 
will help answer some of the questions raised in this paper and further develop this new and exciting field. 

\section{Acknowledgments}
The authors would like to thank Steve Snowden, Francisco Carrera, Mark Devlin 
and Lyman Page for discussion and useful comments. 
This work would have not been possible without the previous work and efforts of others. 
In particular, we have used the following software packages: {\small HEALPIX} (Gorski et al. 2005) and
XSPEC (Arnaud 1996).  We have also made heavy use of the data from BCS (Ebeling et al. 1998), eBCS (Ebeling et al. 2000), 
REFLEX ( B\"ohringer et al. 2004), WMAP (Hinsaw et al. 2009), Diffuse X-ray maps from RASS (Snowden et al. 1996), 
NVSS (Condon et al 1998), QMW IRAS galaxy catalogue (Rowan-Robinson et al. 1991) and BSC (Voges et al. 1999). 
JMD acknowledges support for a Ministerio de Ciencia e Innovaci\'on grant JC2008-00104 from the 
Jose Castillejo program. JMD would like to thank Haverford College for its hospitality during the 
duration of this project. BP's research is supported in part by NSF grant AST-0606975 to Haverford College.


\bsp  
\label{lastpage}  
\end{document}